\documentclass[page-classic]{epl2} 

\title{Evidence for a disorder driven phase transition in 
the condensation of $^4$He in aerogels}

\author{F. Bonnet \inst{1}\and  T. Lambert\inst{1}
\and B. Cross\inst{1} \and L. Guyon\inst{1} \and F. Despetis\inst{2} \and
L. Puech\inst{1} \and P.E. Wolf\inst{1}  }
\shortauthor{F. Bonnet \etal}

\institute{                    
  \inst{1}Institut N\'eel, CNRS/UJF, BP 166,
38042 Grenoble-Cedex 9, France\\
  \inst{2} Groupe d'Etude des Semiconducteurs, CNRS-UMR 5650, Universit\'e Montpellier II
Case Courrier 074, 34095 Montpellier-Cedex 5, France 
}

\abstract{ We report on thermodynamic and optical measurements of the
condensation process of $^4$He in three silica aerogels of different microstructures.
For the two base-catalysed aerogels, the temperature dependence
of the shape of adsorption isotherms and of the morphology of the
condensation process show evidence of a disorder driven transition, in
agreement with recent theoretical predictions.  This transition is not
observed for a neutral-catalysed aerogel, which we interpret as due to
a larger disorder in this case.  }

\begin{document}

\maketitle

\section{Introduction}
The liquid-gas transition of a bulk fluid is an archetypal example of
a first order transition.  Condensation and evaporation of fluids in
various porous media have thus been widely studied in order to
understand the influence of confinement on such transitions.  When the
pressure is increased at a constant temperature, the general picture,
in the case where the liquid phase wets the substrate, is a
progressive adsorption of a thin dense film at low vapour pressure,
followed by a rather abrupt filling below the saturated vapour
pressure.  This phenomenon is ascribed to capillary effects and is
denominated capillary condensation.  In the classical picture, the
difference between the filling pressure and the bulk saturation
pressure decreases with increasing pore size, and the range of
pressures over which the condensation takes place reflects the pores
size distribution.  A characteristic of the capillary condensed regime
is the hysteresis between condensation and evaporation.  The porous
medium empties at a lower pressure than it fills, resulting in an
hysteresis loop in the plane pressure-(average) fluid density.  This
behaviour can have different origins.  In the case of disconnected
pores (as in controlled-pore glasses, e.g. MCM41 silica), it can be
interpreted in terms of the change of the meniscus shape between
filling and emptying\cite{Gelb99}.  For a network of connected pores
of different sizes (e.g., in Vycor silica glass), it is believed to
stem from the energy barrier for nucleating a liquid vapour interface.
A pore cannot empty as long as its smaller neighbours remain filled,
so that emptying involves a specific percolation
process\cite{Mason88}.

Recent numerical studies, based on a mean-field density functional
theory, shed a new light on capillary condensation\cite{Kierlik01,
Sarkisov02}.  A central point is that hysteresis results from the
energetic and geometric disorder of the porous media.  As in
disordered magnetic systems\cite{Sethna}, disorder induces a complex
free-energy landscape, resulting in a large number of metastable
equilibrium states, corresponding to different configurations of the
microscopic liquid-vapour interface.  If thermal fluctuations are
negligible, the system remains trapped in one minimum as the pressure
is increased (or decreased), up to the point where this minimum
disappears and the system moves to another minimum, corresponding to
an avalanche where some region of the porous material transits from
vapour to liquid (or reversely).  The state of the system then depends
on its history, whence hysteresis.

These studies predict the existence of a disorder-driven transition
similar to that occurring in the Random Field Ising Model at zero
temperature \cite{Sethna}.  At small porosity $\mathcal{P}$ (or large
disorder), filling takes place by a succession of small, microscopic
avalanches.  The maximal size of the avalanches increases with
$\mathcal{P}$, up to some critical value $\mathcal{P}_{c}$
(corresponding to some critical disorder), where it diverges.  Above
$\mathcal{P}_{c}(T)$, which increases with increasing temperature $T$,
filling involves a macroscopic avalanche, associated with a jump in
the average fluid density, at some well defined pressure.  This
out-of-equilibrium phase transition implies a change of shape, from
smooth to steep, of the condensation branch of the hysteresis loop,
when the porosity is increased at a constant temperature or,
alternatively, when the temperature is decreased below some critical
value, at a constant porosity\cite{DetcheverryE03}.  In magnetic
systems, an equivalent change of the hysteresis loop has been observed
in Co/CoO disordered films \cite{Berger00}.

Such a change as a function of temperature is not observed in Vycor
and other usual porous materials.  In the new theoretical approach, it
implies that the disorder in these materials is larger than its
critical value at zero temperature.  Thus, a clear demonstration of
the specific role of disorder requires to find a material where the
disorder is small enough to observe the out-of-equilibrium phase
transition at a finite temperature.  Ideal candidates are silica
aerogels, where the silica forms a complex arrangement of
interconnected strands.  Their porosity is large and can be varied in
a wide range (up to 99\%, and more), offering an experimental,
free-standing, realisation of weak and tunable disorder.  Furthermore,
most of the recent theoretical studies have been performed on aerogels
numerically synthesised by Diffusion Limited Cluster Aggregation,
which mimics real aerogels synthesised under basic conditions.  These
numerical studies predict that, for 87\% porosity aerogels, avalanches
are always microscopic, while, for 95\% porosity aerogels, a
transition from microscopic to macroscopic avalanches takes place at a
finite temperature.

Earlier experiments on condensation of helium inside aerogels offer
indeed some support for the new scenario.  Excepting the first
ones\cite{Wong90}, experiments carried out by different groups on
similar aerogels (95\% porosity, gelation process with basic pH), show
that, as in other porous materials, sorption isotherms are hysteretic,
and that condensation occurs over a finite range of pressures
\cite{Tulimieri99,GabayJLTP00,Lambert04,HermanPRB05}.  Still, this
range seems unexpectedly narrow, considering the disordered structure
of the silica strands, especially at large porosity (i.e. weak
disorder) \cite{Tulimieri99,HermanPRB05} or low temperature
\cite{Lambert04, HermanPRB05}.  Steep isotherms have also been
measured using nitrogen as a fluid, but, here, the situation is made
complex by the large contraction of the compliant gels upon
adsorption\cite{Reichenauer01}.  This problem does not arise with
helium, thanks to its much lower surface tension\cite{HermanPRB06}.

The aim of the experiments reported here was to further investigate
the adequacy of the disorder-driven scenario, using combined optical
and thermodynamic measurements to study the effect of temperature on
condensation in three aerogels of different microstructures and
porosities.
\section{Experimental}
Studied aerogels are B100, N102, and B55, where the letter corresponds
to the pH during synthesis (basic or neutral), and the number to the
density (in kg/m$^3$).  We synthesised B100 and N102 (of porosities
$\mathcal{P}$=95.5\%) using a one step process
\cite{PhalippouHandbook}, while B55 ($\mathcal{P}=$97.5\%) has been
synthesised at NorthWestern University by J. Pollanen using a two-step
process.  Aerogels have a fractal structure (characterised by a
fractal dimension $D_{f}$) between the size $a$ of building silica
units and a correlation length $\xi_{G}$.  From previous neutron and
X-ray scattering experiments on similar samples, we expect that
$D_{f}$ and $\xi_{G}$ are similar for B100\cite{Hasmy94,Wang96} and
B55\cite{Pollanen06} (1.7 and 10~nm $\pm$ 20\%), and larger for
N102\cite{Vacher88} (2.6 and 20~nm $\pm$ 20\%).  Since
$\mathcal{P}\propto ({\xi_{G}}/{a})^{3-D_{f}}$, both N102 and B55 must
have a fractal range ${\xi_{G}}/{a}$ larger than B100.  We directly
measured $\xi_{G}$ for our samples using light scattering.  This gives
the structure factor $S(q)$ at zero $q$, which depends on
$\mathcal{P}$, $D_{f}$, and $\xi_{G}$\cite{Ferri91}.  The measured
$S(0)$ corresponds to a light mean free path of 3.5~mm for N102, 50~mm
for B100 and 100~mm for B55.  Using the above $D_{f}$ and
$\mathcal{P}$ values, this yields $\xi_{G}$ values within 20\% of
those expected from the SANS and SAXS measurements.  In particular,
the stronger scattering for N102 reflects its larger $\xi_{G}$.

Aerogels were grown as cylinders (14 mm in diameter for B100 and N102,
10 mm for B55), from which we sliced thin disks (2.7 mm thick for
B100, 4~mm thick for N102 and B55) to fit into a 4~mm thick copper
experimental cell closed by two sapphire windows.  The cell was
mounted in a cryostat with 8 optical ports 45 $^{\circ}$ apart,
allowing observation of samples under different scattering angles.
Sorption isotherms were performed on these samples between 4.2~K and
the critical bulk temperature.  The pressure in the cell was measured
as a function of the injected helium amount, and simultaneous light
scattering measurements were performed to probe the distribution of
helium inside the aerogels.  Except for N102 (due to its strong
intrinsic scattering), the level of multiple scattering was generally
weak.  This is a specific advantage of the low index of refraction of
helium (1.024 at 4.2~K).

The flow in or out from the cell was controlled with  a regulated 
flowmeter for B100 and N102, and by
changing the temperature of an external reservoir connected to the
cell for B55\cite{Cross07}. The flowrate must be small enough that the heat
released by the condensation process creates a negligible temperature
gradient inside the aerogel.  For B100 and N102, the filling time 
corresponding to the hysteretic part 
of the adsorption isotherm was between 15 to 30 hours, which we
checked to be long enough.  For B55, it ranged from 6 to 
48 hours.

The helium mass in the aerogel is obtained by integrating the flowrate
over time (for B100 and N102) or from the temperature of the reservoir
(for B55), and correcting for cold and warm dead
volumes\cite{Lambert04}.  Conversion of this mass into a condensed
fraction $\Phi$ must take into account the compression of the helium
close to the silica, especially for temperatures close to the bulk
critical point.  We thus impose $\Phi$=1 for the filled aerogel (i.e.
zero vapour fraction) and compute the change of $\Phi$ from the change
in mass from this initial point, assuming that the density of the
dense phase and of the vapour inside the aerogel both have their bulk
value\cite{Lambert04}.  The result will be correct as long as the
first layers of helium are not affected, i.e. except for small
fractions.

\section{Results}
\begin{figure}[t!]
\onefigure[width=8 cm]{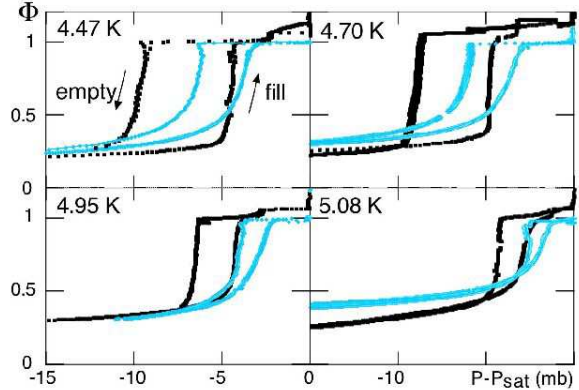}
\caption{ Hysteresis loops at four
temperatures, 4.47, 4.71, 4.96 and 5.08~K, for two aerogels of
porosity 95\%, B100 (black) and N102 (grey).  The liquid fraction
$\Phi$ in aerogel is plotted versus $P-P_{sat}$, where
$P_{sat}$ varies between 1260 and 2090 mb; for B100, there is a clear
change of shape with temperature.}
\label{isothermes}
 \end{figure}
 
\begin{figure}[t!]
\onefigure[width=8 cm]{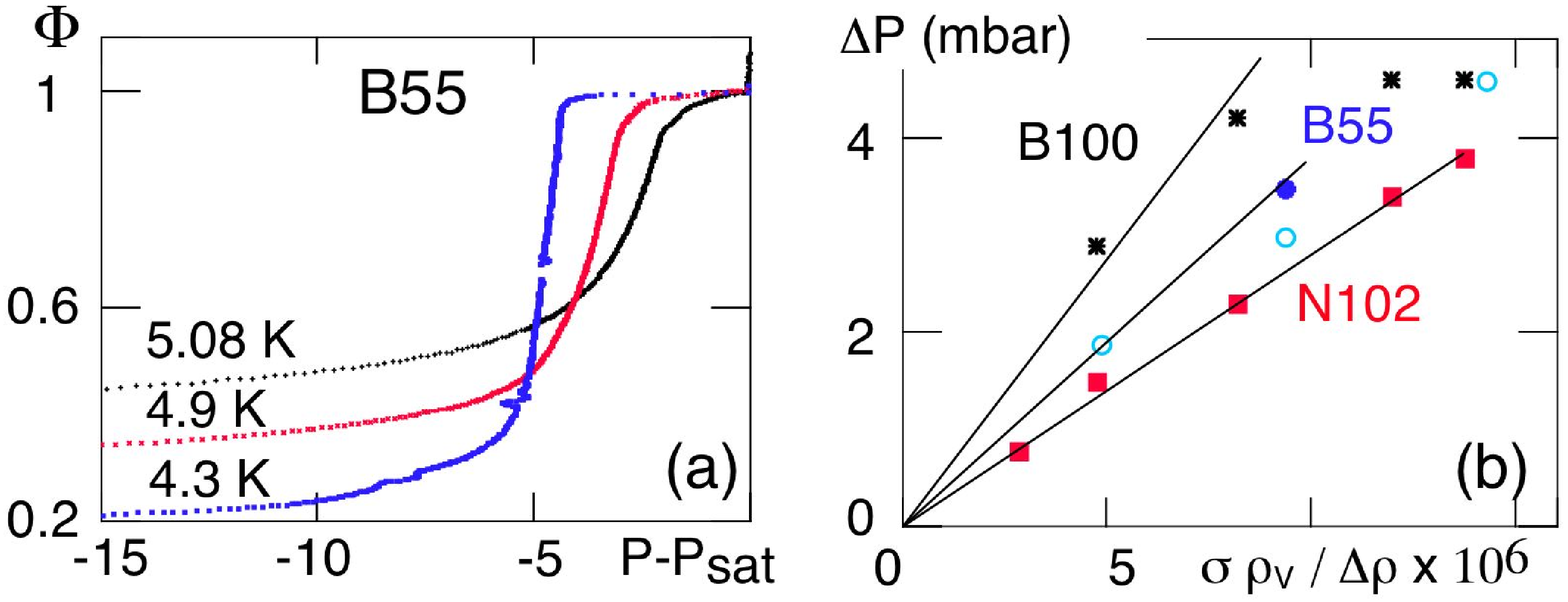} 
\caption{a) B55 raw adsorption
isotherms for a 0.2 STP cc/min flowrate, corresponding to a filling
time of order 12 hours; b) Position $\Delta P$ of the point of the
condensation branch with maximal slope versus the temperature
dependent ratio $\sigma \rho_{V}/(\rho_{L}-\rho_{V})$ (in $J/m^{2}$)
for the different samples.  For B55, the filling time is 12~h for the
open circles and 48~h for the closed one.  The straight lines
correspond to average pore diameters of 18, 25, and 36~nm.}
\label{isothermesB55}
\end{figure}

Isotherms for the three samples are shown in 
figs.~\ref{isothermes} and ~\ref{isothermesB55}a.
Figure~\ref{isothermes} compares the isotherms obtained for B100 and
N102.  The general trends are in qualitative agreement with the usual
picture of capillary condensation; the hysteresis loop gets narrower
and closer to the bulk saturation pressure $P_{sat}$ as the surface
tension decreases with increasing temperature.  Also, it is smoother,
and closer to $P_{sat}$ for N102, consistent with the wider
distribution of pores extending up to larger sizes.

We quantify the position of the adsorption isotherm by $\Delta
P=P_{max}-P_{sat}$, where $P_{max}$ is the pressure where its slope is
the largest.  For a cylindrical pore of radius $R$, filling occurs at
the stability limit of the adsorbed film.  For $R$ much larger than
the film thickness, this gives \cite{Everett72,SaamColePRB75} $\Delta
P = \rho_{V}/(\rho_{L}-\rho_{V}) .  \sigma/R $, where $\sigma$ is the
helium surface tension, and $\rho_{L}$ and $\rho_{V}$ are the bulk
liquid and vapour densities at saturation.  For a collection of
independent pores, $\Delta P$ corresponds to the average pore size.
Although, \it a priori \rm, aerogels cannot be so simply described,
fig.~\ref{isothermesB55}b shows that the equation above accounts for
the temperature dependence of $\Delta P$ for N102 in the whole
temperature range, and for B100 above 4.95~K, with characteristic
`pores' sizes $R\simeq$ 35~nm and 20~nm.  Both values are about twice
the measured correlation lengths.  For the larger porosity sample B55,
$\Delta P$ is affected by the flowrate used, due to the heating effect
of adsorption, but the error is small, as shown by the point for a
slow filling at 4.89~K. We estimate the corresponding average pore
size to be 25~nm, intermediate between B100 and N102.  Comparison of
B100 and N102 show that the average pore size at constant porosity
increases with the gel correlation length $\xi_{G}$, while comparison
of B100 and B55 shows that, at given $\xi_{G}$, it depends on
porosity.  In other words, both the porosity and the correlation
length influence the position of the isotherms.

\begin{figure}[t!]
\onefigure[width=8 cm]{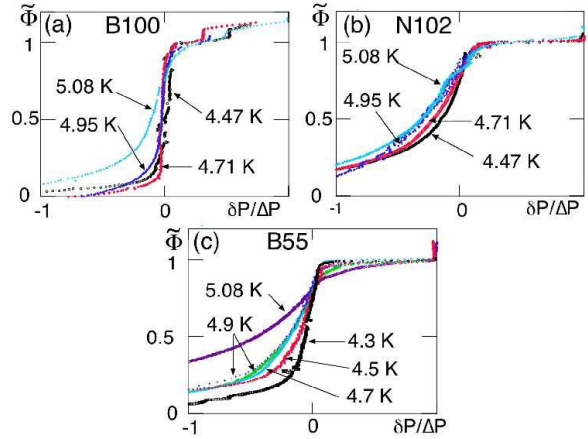} 
\caption{Normalised
isotherms for B100 (a), N102 (b), and B55 (c).  The abscissa is
$\delta P=P-P_{max}$ normalized by $\Delta P$, which equals 1 at
$P_{sat}$.  $\tilde{\Phi}$ is the condensed fraction excluding the
adsorbed film (see text).  For B55, flowrates range between 0.05 and
0.4~STPcc/min.  Although $\Delta P$ at 4.9~K depends on the flowrate,
the normalized isotherm does not.  The temperature dependence for B100
is consistent with a disorder driven transition occurring occurring
between 4.71~K and 4.95~K.}
\label{isosnorm}
\end{figure}
For both B100 and B55, the isotherms are steeper at low than at large
temperatures.  This is inconsistent with the usual picture of
capillary condensation, where the shape of the isotherms reflects the 
pore distribution.

This is not an effect of adsorption.  For a flat substrate, the film
thickness at $\Delta P$, being proportional to
$((\rho_{L}-\rho_{V})/{\sigma})^{1/3}$, increases with temperature.
This also holds for a cylindrical pore\cite{SaamColePRB75}.  We thus
expect that the fraction $\Phi_{0}$ at the lower closure point of the
hysteresis loop increases with temperature, which is in agreement with
fig.~\ref{isothermes}.  In order to take into account this effect when
comparing the isotherms between different temperatures, we transform
$\Phi$ to $\tilde{\Phi}$=$(\Phi-\Phi_{0})/(1-\Phi_{0})$, the condensed
fraction excluding the film.  We also normalise the pressure axis by
$\Delta P$, to account for the different average pores sizes and the
temperature dependence of $\sigma$ and $\rho_{L}-\rho_{V}$.
Figure~\ref{isosnorm} shows that the thus normalized isotherms
collapse for N102.  This is consistent with the usual picture of
capillary condensation, where the spatial distribution of liquid
(hence the interface curvature, to which the pressure is related
through Kelvin's equation), only depends on the condensed fraction,
not on the temperature.  In contrast, such a picture breaks down for
the base-synthetised samples, for which the isotherms become steeper
at low temperature.  The behaviour for these samples supports the
scenario of a disorder-driven transition.  For B100, the normalized
isotherm is nearly vertical for 4.47~K and 4.71~K, implying that the
critical temperature lies between 4.71~K and 4.95~K. Although there is
no corresponding theoretical prediction, this transition might also
cause the anomalous behaviour of $\Delta P$ below 4.95~K. For B55, the
isotherms are less steep than for B100, and the effect of temperature
is not as pronounced.  This is surprising as measurements by Herman
{\it{et al}\rm} show a much steeper isotherm for a 97.5\% sample than
for a 95\% sample, both one-step synthesised.  The smoother isotherms
for our B55 could result either from a broadening caused by large
heterogeneities of the average silica density, or from the different
microstructure induced by the two-step synthesis.

\begin{figure}[h]
\onefigure[width=8 cm]{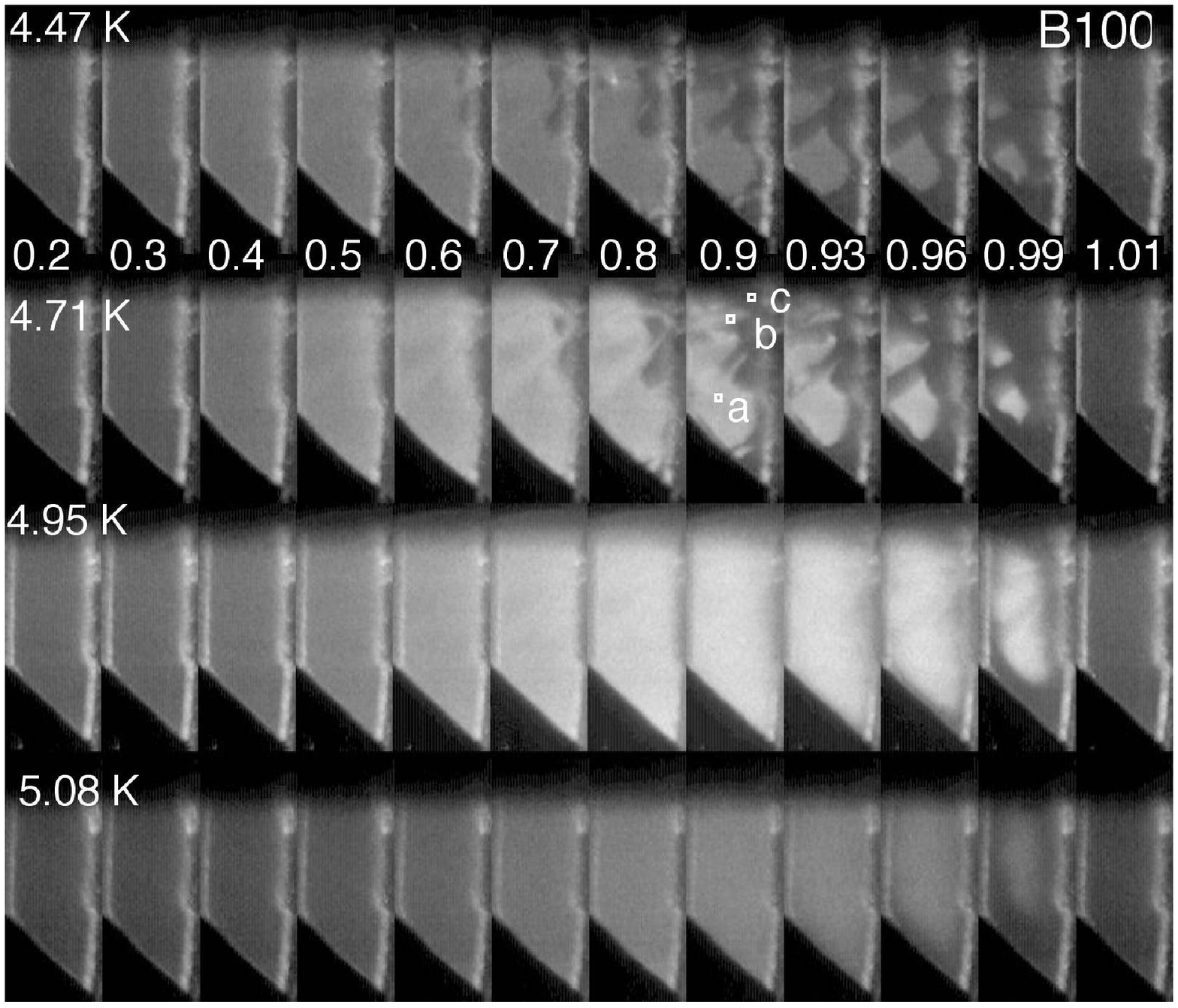} 
\onefigure[width=8
cm]{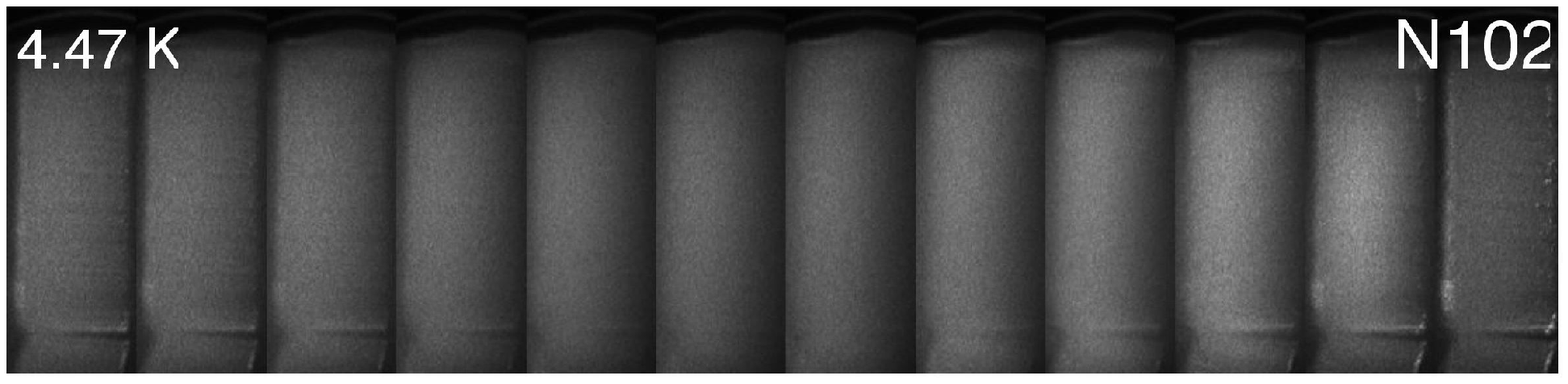} 
\onefigure[width=8 cm]{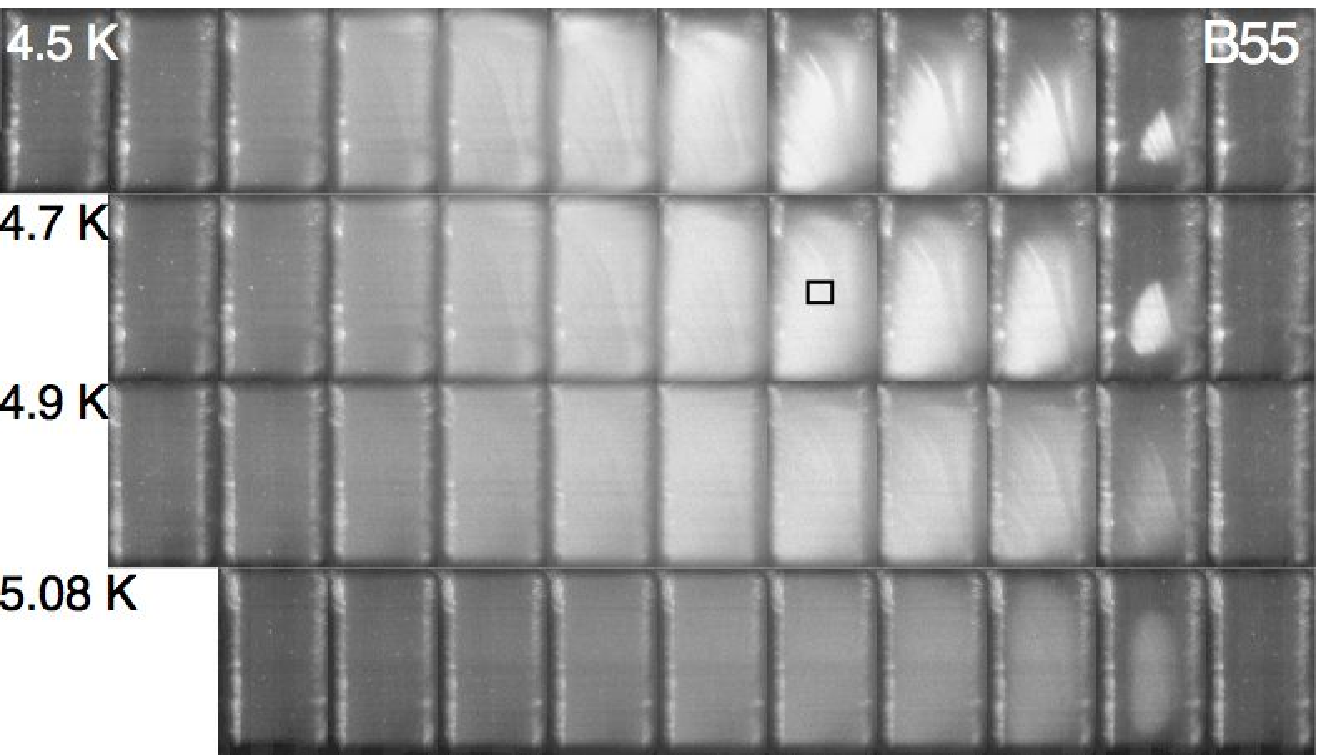} 
\caption{
Images observed at 45$^\circ$ from the incident laser sheet for
condensed fractions from 0.2 to 0.9 (by steps of 0.1), 0.93, 0.96,
0.99 and 1.01.  The width of the images is 4~mm; top panel : B100
(logarithmic grey scale); middle panel : N102 at 4.47~K (linear grey
scale); bottom panel : B55 (logarithmic gray scale).  The rectangles
for B100 and B55 are those used to measure the intensity in
figs.~\ref{imageB} and \ref{imageH}.}
\label{photos}
\end{figure}
\begin{figure}[t!]
\onefigure[width=8 cm]{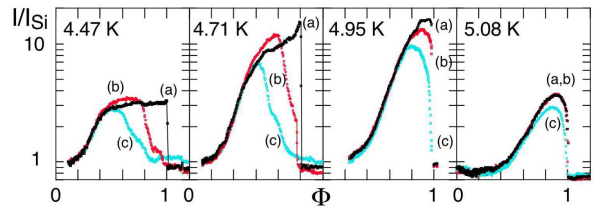}
\caption{ Scattered intensity for B100, referred to the empty aerogel 
situation, for the three regions labelled (a), (b), (c) in
fig.~\ref{photos}a.  The smaller scattering at low temperature points
to a temperature dependence of the microscopic distribution of liquid
at a given condensed fraction $\Phi$.}
\label{imageB}
\end{figure}

We now come to the optical measurements.  We illuminate the aerogel by
a thin (100 $\mu$m wide), vertically polarised, He-Ne laser sheet
under a 45$^{\circ}$ incidence with respect to its faces, and image it
at 45$^\circ$, 90$^\circ$, and (for N102 and B55) 135$^\circ$ using
CCD cameras.  At any point of the intercept of the aerogel by the
laser sheet, the brightness of the image is proportional to the
intensity scattered in the direction of observation.  The upper part
of fig.~\ref{photos} shows images of B100 for increasing helium
fractions along the adsorption isotherms, and fig.~\ref{imageB} the
scattered intensity by the three selected spots in the images.
Quantitative analysis of these curves show that, below
$\Phi\approx$0.3, the scattered intensity is that expected for a thin
film of helium covering uniformly the silica strands, whereas, beyond
this value, it becomes much larger, implying the formation of liquid
clusters correlated over distances larger than the correlation length
of the silica\cite{Lambert04}.
Strikingly, while this increase of the scattered intensity occurs
uniformly over the sample for the two larger temperatures, for the two
lower ones, it is only so for fractions less than about 0.5.  Beyond,
liquid progressively invades the aerogel, resulting in the growth of
dark regions.  Since the macroscopic morphology is identical for
4.47~K and 4.71~K, and for different rates of condensation as well, it
must reflect macroscopic, heterogeneities of the average density of
silica, which make some regions favour slightly more the liquid.  The
fact that the filling pattern is homogeneous above 4.95~K then implies
that the isotherm intrinsic width is larger than its broadening due to
the heterogeneities, the reverse being true below 4.95~K. The observed
change of morphology is thus a signature of the change of slope of the
isotherm.

These observations exhibit the behaviour expected for macroscopic
avalanches; at any given spot, the transition to the dark state
occurs discontinuously, when this spot is swept by the boundary
between the bright and the dark regions.  This is consistent with the
occurrence of a macroscopic (but local) avalanche from a partly filled
to a fully filled state.  This will make the pressure at which the
avalanche takes place to vary continuously in space, so that the
transition should occur successively at different locations in the
aerogel, in qualitative agreement with the observations.  
\begin{figure}[t]
\onefigure[width=8 cm]{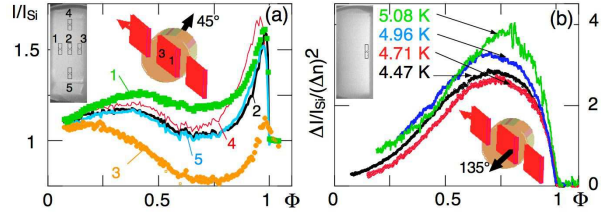}
\caption{ Intensity scattered by N102 as a function of the condensed
fraction $\Phi$ : a) at 4.47~K and 45$^\circ$.  The signal,
normalized by the silica contribution, corresponds to five regions at
three different depths inside the aerogel disk.  Differences between
curves are consistent with
multiple scattering effects combined with an homogeneous distribution
of the liquid clusters within the sample, in contrast to B100 at the
same temperature; b) at 135$^\circ$, close to the entrance of the
laser sheet.  The silica contribution $I_{Si}$ has been subtracted,
and the result normalized by $I_{Si}$ and $({\Delta n 
(T)}/{\Delta n (4.47)})^2$, the square of the optical contrast $\Delta
n$ normalized to its value at 4.47~K. The signal is similar at all
temperatures, suggesting that, unlike for B100, the spatial
distribution of the liquid does not depend on temperature.}
\label{imageN}
\end{figure}

Thus, the results for B100 lead us to suggest that this sample
exhibits a disorder induced transition, which the change of filling
pattern would be another marker of.  The relationship between the filling 
pattern and the shape of the isotherms is supported by the behaviour
observed for N102 at 4.47~K in fig.~\ref{photos}.  Due to the strong
scattering of the aerogel, the images are blurred, but do not show any
sign of heterogeneity except very close to filling, consistent with
the fact that the isotherm is smooth even at this temperature.  The
homogeneity of the filling pattern is confirmed by the quantitative
analysis of fig.~\ref{imageN}(a)~:~at any fraction $\Phi$,
the intensity scattered at 45$^\circ$ is the same along any vertical
line of the picture, corresponding to a given depth, hence path
length, inside the sample.  There is a difference between points at
different depths, but this (as well as the bump in the signal at $\Phi
\approx$ 0.4) can be explained by multiple scattering effects.

For B55, the behaviour resembles that observed for B100.
Figure~\ref{photos} shows heterogeneities in the form of filaments
above $\Phi\approx$ 0.5, which disappear around 4.89~K. This change of
morphology, together with the isotherms of fig.~\ref{isothermesB55},
suggests that a disorder driven transition also occurs for this
sample.  If this is true, the critical temperature would not be very
different from that for B100, despite the larger porosity.  This, as
well as the absence of transition for N102, shows that the 
pertinent parameter for determining the disorder-induced transition 
involves not only the porosity, but the whole
microstructure. 

Another way of demonstrating the disorder driven scenario would be to
measure the microscopic distribution of liquid as a function of
condensed fraction $\Phi$.  Our experiments already allow
to compare the size of the correlated helium clusters (droplets or
bubbles) between different temperatures and samples.  Assuming
independent and spherical clusters, the scattered intensity depends on
their size and density, or, equivalently, on their size and local
liquid (or vapour) volume fraction.  For sizes smaller than the
wavelength, the Rayleigh-Gans scattering regime applies, and, at a
given fraction, the scattered intensity is expected to increase with
the size and to be proportional to the contrast factor $(\Delta
n)^{2}$, where $\Delta n$ is the difference in refractive index
between the liquid and the vapour.
\begin{figure}[t]
\onefigure[width=8 cm]{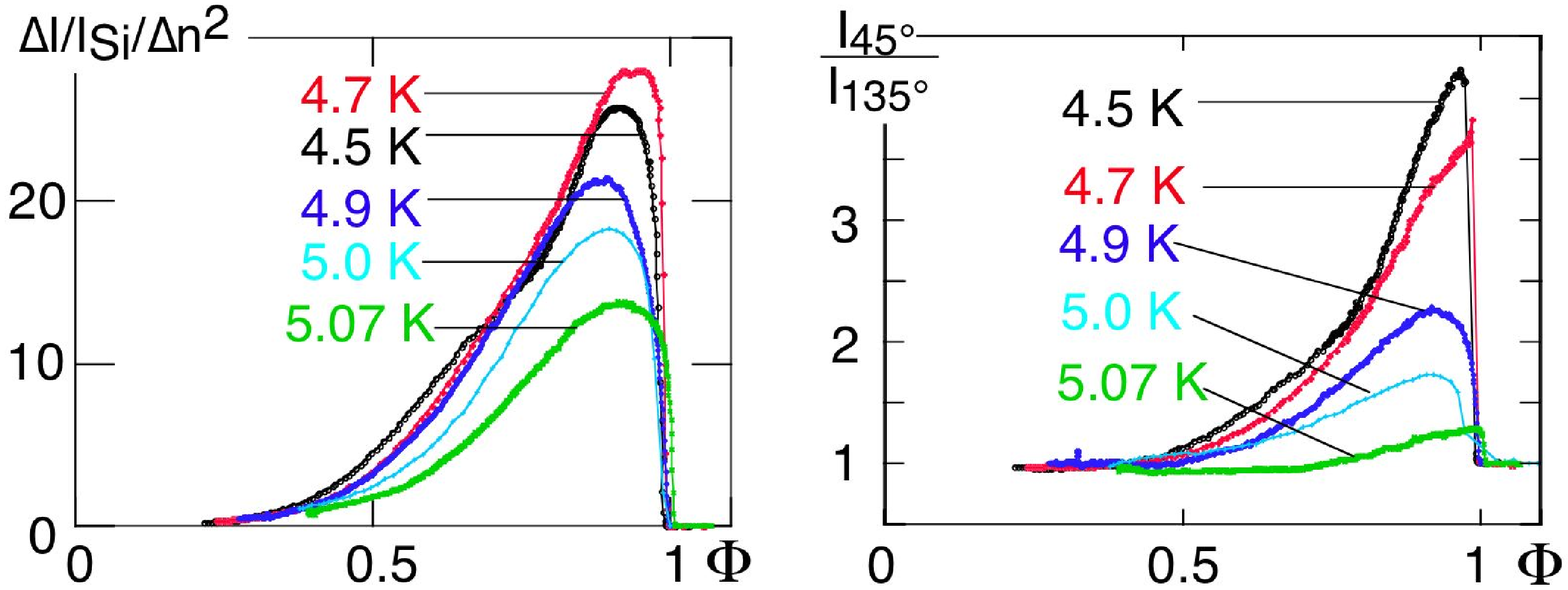} \caption{ a) Normalized
intensity scattered at 45$^\circ$ by B55 as a function of the
condensed fraction $\Phi$. The studied region is the rectangle in 
fig.~\ref{photos}.
The decrease of signal with increasing
temperature indicates that the size of correlated domains decreases.
b) Anisotropy of scattering between 45$^\circ$ and 135$^\circ$.  Its
$\Phi$ and temperature dependencies show that the correlated domains
size reaches a maximum close to filling, and  that the maximal size 
is larger than 100~nm and decreases
with increasing temperature.}
\label{imageH}
\end{figure}

Comparison of figs.~\ref{imageB},~\ref{imageN}b, and \ref{imageH}a
shows different behaviours for the different samples.  For B100, the
scattered intensity, at a given $\Phi$, is smaller at 4.46~K than at
4.71~K (which is obvious on the images of fig.~\ref{photos}), although
the contrast factor is larger.  In the low temperature regime, the
size of liquid clusters thus increases with temperature.  In contrast,
for N102, once rescaled by the contrast factor, the intensity
scattered at 135$^\circ$ close to the entrance of the laser
sheet\footnote{At other positions inside the aerogel, or at
45$^\circ$, the attenuation due to scattering along the light path
does not allow such a direct comparison.} behaves roughly the same,
whatever the temperature.  This suggests that, at a given $\Phi$, the
distribution of liquid does not depend too much on temperature.  For
B55, the signal is temperature dependent, but does not decrease at low
temperature.  The low temperature behaviour of B100 thus appears
singular, and its origin remains to be understood.

For all samples, the optical signal reaches its maximum between
$\Phi$=0.75 and 0.95.  In this range, it probably originates from
bubbles of vapour.  Assuming these bubbles to be spherical, their size
can be estimated from the scattered intensity (for the case of
homogeneous filling, when the local liquid volume fraction coincides
with the measured global fraction $\Phi$), or from the anisotropy of
scattering (fig.~\ref{imageH}b).  Both estimates give a diameter of
order 200 to 300~nm for B100 and B55, which tends to decrease with
increasing temperature above 4.9~K. For N102, the sharp peak at
45$^\circ$ in fig.~\ref{imageN}a also implies a large diameter.  Due
to the complexity brought by multiple scattering, only a lower bound
of about 200~nm can be determined.

Numerical studies of the avalanches size distribution for basic-like
aerogels of porosity between 87\% and 95\% have shown that the
avalanches size increases along the adsorption isotherms, to reach a
maximum close to filling, and also increases with increasing porosity
or decreasing temperature\cite{DetcheverryE05}.  These predictions
cannot be directly compared to our experiments, as we only measure the
size of correlated liquid or vapour domains.  However, close to
$\Phi$=1, one might expect that the avalanches consist in filling the
last surviving bubbles.  The size decrease of these bubbles with
increasing temperature would be qualitatively consistent with the
theory.  Quantitatively, however, numerical studies predict a size of
1 to 2~$\xi_{G}$, depending on temperature, whereas we find bubbles
sizes larger than 10~$\xi_{G}$.  A more direct comparison to theory is
possible by using the results of a numerical study on a 87\% porosity
aerogel at low temperature, the behaviour of which is expected to be
similar to a 95\% porosity aerogel above its transition
temperature\cite{DetcheverryE06}.  This study predicts that the range
of fluid-fluid correlations reaches a maximum of order 2~$\xi_{G}$
close the point of maximal slope in the isotherm.  While the position
of this maximum is consistent with our experiments, its value is again
lower than the measured size of correlated regions.  Further work,
both theoretical and experimental, is needed to understand this
discrepancy.

\section{Conclusions}
We have shown that different light aerogels behave very differently
with respect to adsorption.  Our results provide a strong evidence for
the occurrence of a disorder induced transition as a function of
temperature in B100.  At the same porosity, N102 does not present any
transition.  In the framework of the disorder-driven transition, this
implies a larger disorder in this case, which would come from the
wider distribution of length scales for this aerogel.  It would be
interesting to perform numerical studies on similar aerogels to
support this interpretation.  In the lighter aerogel B55, there seems
to be a transition, but not as clear as for B100.  This contrasts with
the theoretical expectations and could result either from a less
homogeneous sample or from the different microstructure due to the
method of synthesis.  Checking whether the transition moves to larger
temperatures when the porosity increases will therefore require to
compare aerogels of different porosities synthesised by the same
method.  Finally, our experiment demonstrates that optical
measurements are a powerful tool, complementary to the isotherms, for
the study of adsorption of helium in aerogels, and more generally, in
porous media.  
\acknowledgments
We acknowledge T. Herman, J. Beamish,
F. Detcheverry, E. Kierlik, M.L. Rosinberg, G. Tarjus, and J.
Phalippou for useful suggestions or discussions.  We are grateful to
J. Pollanen and W. P. Halperin for providing sample B55 with support
from NSF DMR-0703656.  Studies on B55 were supported by
ANR-06-BLAN-0098.  

\end{document}